# A 30nA Quiescent 80nW to 14mW Power Range Shock-Optimized SECE-based Piezoelectric Harvesting Interface with 420% Harvested Energy Improvement

Anthony Quelen, Adrien Morel, Pierre Gasnier, Romain Grézaud, Stéphane Monfray, Gaël Pillonnet

Univ. Grenoble Alpes, CEA, LETI, MINATEC, F-38000 Grenoble, France.
STMicroelectronics, Grenoble, France.

Piezoelectric Energy Harvesters (PEH) are usually used to convert mechanical energy (vibration, shocks) into electrical energy, in order to supply energy-autonomous sensor nodes in industrial, biomedical or domotic applications. Non-linear extraction strategies such as Synchronous Electrical Charge Extraction (SECE) [1-2], energy investing [3] or Synchronized Switch Harvesting on Inductor (SSHI) [4] have been developed to maximize the extracted energy from harmonic excitations. However, in most of today's applications, vibrations are not periodic and mechanical shocks occur at unpredictable rates [4]. SSHI interfaces naturally seemed to be the most appropriate candidate for harvesting shocks as they exhibit outstanding performance in periodic excitations [4]. However, the SSHI strategy presents inherent weaknesses while harvesting shocks, since the invested energy stored in the piezoelectric capacitance cannot be recovered.

In this work, we propose a self-starting, battery-less, 0.55mm$^2$ integrated energy harvesting interface based on SECE strategy which has been optimized to work under shock stimulus. Due to the sporadic nature of mechanical shocks which imply long periods of inactivity and brief energy peaks, the interface's average consumption is optimized by minimizing the quiescent power. A dedicated energy saving sequencing has thus been designed, reducing the static current to 30nA and enabling energy to be extracted with only one single 8μJ shock occurring every 100s. Our SECE-based circuit features a shock FoM 1.6x greater than previous SSHI-based interfaces [4].

The proposed system depicted in Fig.1 is made of a negative voltage converter rectifying the PEH output voltage, and a SECE power path controlled by a sequenced circuit. The sequencing is divided in 4 phases and the associated time diagrams are illustrated in Fig.2. During the sleeping mode $T_1$, all blocks except the shock detection (SD) are turned off. During the starting phase, the energy is stored in $C_{ASIC}$ through a cold-start path, increasing $V_{ASIC}$. This will progressively turn on the SD. Next, when stress applied to the piezoelectric material leads to an increase in $V_{REC}$, the SD checks if the electrical energy converted by the piezoelectric transducer is sufficiently high to be harvested ($V_{REC}>V_{ASIC}$). By setting $Flag_{SHOCK}$, the SD enables the $V_{ASIC}$ detection which determines whether the cold start path should be activated. If $V_{ASIC}$ is below 1.5V, we consider that the stored energy is insufficient to start the SECE operation, and the cold start path remains connected in order to keep on charging $C_{ASIC}$. If this is not the case, the detection block sends the $Flag_{START}$ signal which disables the cold start, enables the peak detection and starts the maximum voltage detection phase $T_2$. When $V_{REC}$ reaches its maximum, the system enters its harvesting phase $T_3$. $V_{N1}$ is set high, which connects the inductance L with the piezoelectric capacitance, $C_P$. The dual mode comparator (DMC) is used in its zero crossing detection (ZCD) configuration, and detects when $V_{REC}$ goes below $V_{TL}$=-14mV, which means that almost all the energy previously stored in $C_P$ has been extracted in L. Then, the system starts its storing phase $T_4$. $N_1$ is turned off, while $P_2$ and $N_2$ are turned on. The instant $I_L$ reaches zero, which is detected by the same DMC used in its Reverse Current Detection (RCD) configuration, indicates that all the energy that was stored in L during $T_3$ has been transferred in $C_{STORE}$. Ultimately, the system returns to its sleep mode $T_1$, waiting the next energy event. $N_3$ acts as a freewheeling diode and provides a path to dissipate the remaining energy in L.

Fig.3 shows detailed transistor-level schematics of the DMC and the Shock and $V_{ASIC}$ Detections. During $T_3$, the DMC is enabled in its ZCD configuration. $M_2$ and $M_4$ constitute a differential pair allowing $V_{REC}$ to be compared with the ground voltage. Due to $M_1$, when $V_{REC}$ is high, only ¼ of the bias current flows through $M_1$ and $M_2$. As $V_{REC}$ decreases (thanks to the charge transfer occurring between $C_P$ and L), the current in $M_2$ is increased, which improves the detection accuracy. Furthermore, the circuit consumption is reduced when $V_{REC}$ is high, since it is only useful to increase the comparator performances when $V_{REC}$ gets close to 0V. $V_{HYST}$ is initially high, which creates a -14mV offset on the input of the comparator. The circuit also includes a comparator which is used to accurately implement the zero crossing detection. During $T_4$, the DMC switches to its RCD configuration. $V_{HYST}$ is set low, which suppresses the -14mV offset. In this phase, $V_{REC}$ is proportional to $-I_L$, as $N_2$ is turned on. Therefore, when $I_L$ decreases, $V_{REC}$ increases until it reaches 0V. Then, $T_1$ starts. The DCM is disabled in order to avoid any unnecessary energy consumption, and only the SD is powered. Therefore, during $T_1$, the 30nA@1.5V current drawn from $C_{ASIC}$ is the one flowing through $M_{17}$, as shown in Fig.3. When $V_{REC}$ increases, current starts flowing through $M_{14}$ which forces $M_{16}$'s drain potential to increase. If $V_{REC}>V_{ASIC}$, then $Flag_{Shock}$ becomes high, which consequently enables the $V_{ASIC}$ Detector by forcing $M_{20}$ conduction. To avoid any ringing, a resistance $R_{HYS}$ is used to create a difference between the high (1.5V) and low (1.4V) threshold. The integrated resistances $R_1$ and $R_2$ enable the minimum of $V_{ASIC}$ to be selected, to ensure the self-operation of the chip. In our case, we fixed this minimum $V_{ASIC}$ at 1.5V. When this condition is satisfied, $Flag_{Start}$ is set high, thanks to a two-stage comparator.

Our chip was fabricated in CMOS 40nm technology including 10V devices, and occupies a 0.55mm$^2$ core area (Fig.7). In order to emulate both periodic and shock excitations, a MIDE piezoelectric generator (PPA1011) with a 5.67g mobile mass and a resonant frequency of 75.4 Hz has been placed on a shaker. The harvester has an intrinsic capacitance $C_P$ of 43nF. Fig.4 shows multiscale experimental waveforms of the interface circuit undergoing shocks with $C_{STORE}$ and $C_{ASIC}$ initially discharged. The shocks are applied every second, with various accelerations from 5 to 16G. The off-chip inductance L and capacitances $C_{STORE}$ and $C_{ASIC}$ values are 2.2mH, 100µF, and 10µF, respectively. After the first three shocks, which are used to store enough energy in $C_{ASIC}$ thanks to the cold start power path, the system operates autonomously in its optimized mode and the energy is stored in $C_{STORE}$. For test purposes, when $V_{STORE}$ reaches 2.8V, the energy monitoring block intermittently connects a 1kΩ load resistance to emulate the consumption of a sensor.

The power stored in $C_{STORE}$ was measured under shock and periodic vibrations for many $V_{STORE}$ as shown in Fig.5. From weak to strong shocks, our chip harvested 2.8x to 4.2x more than the maximal energy harvested using an on-chip full bridge rectifier interface, while it reached a FoM of 3.14 under periodic excitation. In Fig.6, the performance of our chip is compared to prior art. We obtained a 1.6x shock FoM enhancement in comparison to previous work [4]. Our system also shows the best FoM under periodic excitation compared to other SECE interfaces [1-2]. The measured maximum end-to-end efficiency of our circuit is 94% under periodic excitation at 82µW which is the highest end-to-end efficiency compared to former work [1-4]. The measured quiescent current in sleeping mode is 30nA@1.5V. This allows self-operation of our circuit with an input power as low as 80nW. We were able (using various PEH) to maintain an efficiency over 70% for input power below 14mW. The proposed IC in 40nm technology allows to add harvesting functionalities within a microcontroller die.

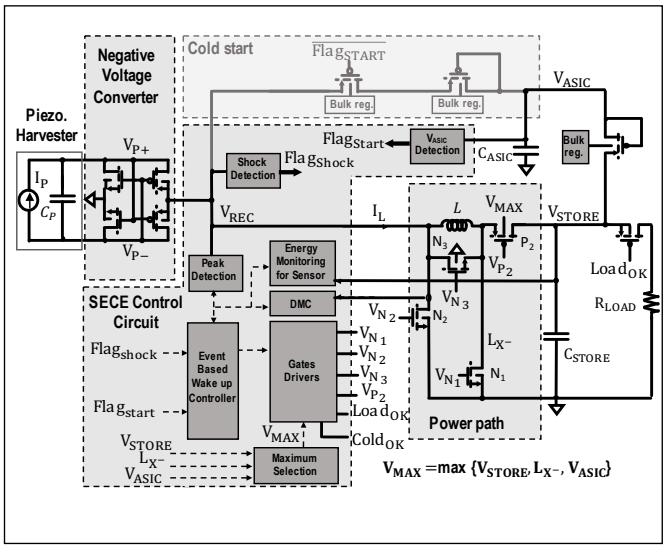

Figure 1: Piezoelectric interface overview.

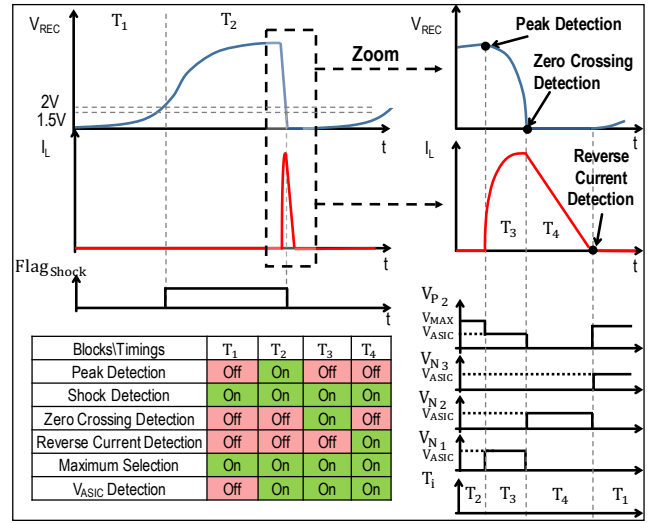

Figure 2: Waveforms, chronograms and sequencing of the control circuit.

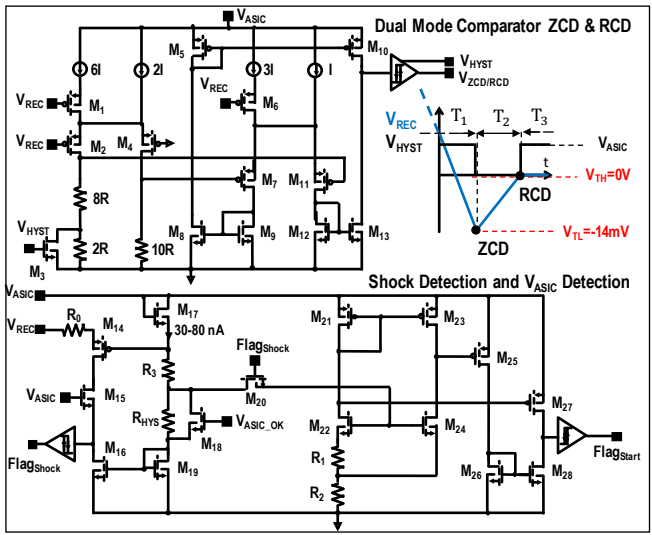

Figure 3: Dual mode comparator and shock detector schematics

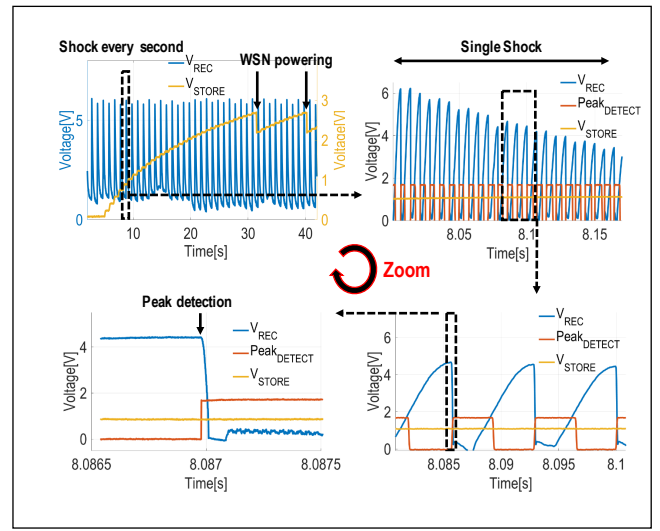

Figure 4: Measured transient waveforms of the proposed interface.

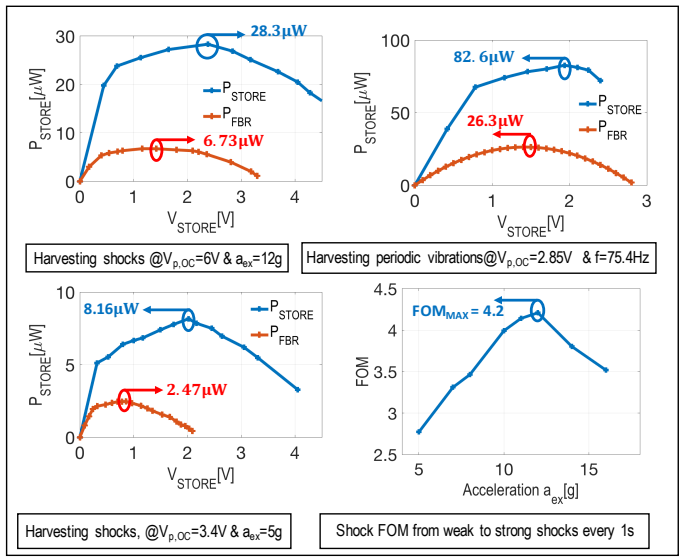

Figure 5: Measured harvested power comparison between our interface and using an active standalone full bridge rectifier (FBR).

Figure 6: Performances comparison with prior art.

|  | [1] | [2] | [3] | [4] | This Work | Unit |
|---|---|---|---|---|---|---|
| Technology | 350 | 350 | 350 | 350 | 40 | nm |
| Chip Size | 3.6 | 1.25 | 2.34 | 0.72 | **0.55** | mm$^2$ |
| Scheme Type | SECE | SECE | Energy Investing | SSHI | SECE | - |
| Piezoelectric Harvester | Murata | MIDE V22B | MIDE V22B | MIDE V21B & V22B | MIDE PPA1011 | - |
| $C_P$ | 23 | 19.5 | 15 | 26 | 43 | nF |
| Excitation type | Periodic | Periodic | Periodic & Shock | Periodic & Shock | Periodic & Shock | - |
| Operation Frequency | 100 | 174 | 143 | 225 | 75.4 | Hz |
| FOM (periodic)[2] | ≈170 [1] | 206 | 360 | **681** | 314 | % |
| FOM (shocks)[3] | N/A | N/A | - | 269 | **420** | % |
| Cold Startup | Yes | Yes | No | Yes | Yes | - |
| End-to-end Efficiency | 61 | 85.3 | 69.2 | ≈88 [1] | 94 | % |
| Input power range | 10-1000 | 5-500 | - | 4-1000 | **0.080-14000** | µW |
| Quiescent current | 1 | 0.3 | 0.1 | ≈1 [1] | 0.03 | µA |

(1) Calculated from the paper   (2) FOM (periodic) = $\frac{\max(P_{out})}{f \cdot V_{oc}^2 \cdot C_p}$   (3) FOM (shocks) = $\frac{\max(P_{out})}{\max(P_{out\ FBD})}$